\documentclass[twocolumn,showpacs,amsmath,amssymb,prl]{revtex4}

\usepackage{graphicx}%
\usepackage{dcolumn}
\usepackage{amsmath}
\newcommand{\be}{\begin{equation}}
\newcommand{\ee}{\end{equation}}
\newcommand{\bey}{\begin{eqnarray}}
\newcommand{\eey}{\end{eqnarray}}
\newcommand{\bw}{\begin{widetext}}
\newcommand{\ew}{\end{widetext}}

\newcommand{\ole}{\overline}

\newcommand{\ra}{\rangle}
\newcommand{\la}{\langle}
\newcommand{\br}{ {\bf r} }
\newcommand{\bp}{ {\bf p} }

\begin{document}
 %\draft

 \title {
 Stability of Quantum Motion: Beyond Fermi-golden-rule
 and Lyapunov decay
 }

 \author{Wen-ge Wang$^{1,2}$, G.~Casati$^{3,4,1}$, and Baowen Li$^1$}

 %\address{
 \affiliation{
 $^1$Department of Physics, National University of Singapore, 117542 Singapore
 \\ $^{2}$Department of Physics, Southeast University, Nanjing 210096, China
 \\ $^{3}$Center for Nonlinear and Complex Systems, Universit\`{a}
 degli Studi dell'Insubria and Istituto Nazionale per la Fisica della Materia,
 Unit\`{a} di Como, Via Valleggio 11, 22100 Como, Italy
 \\ $^4$Istituto Nazionale di Fisica Nuclear, Sezione di Milano, Via Celoria 16, 20133 Milano, Italy
 }

 \date{10 Spetember 2003}

 \begin{abstract}

 We study, analytically and numerically, the stability of quantum motion for a classically chaotic system.
 We show the existence of different regimes of fidelity decay.
 In particular, when the underlying classical dynamics is weakly chaotic,
 deviations from Fermi-Golden-rule and Lyapounov regimes are observed and discussed.

 \end{abstract}


\maketitle

 %\begin{multicols}{2}

 The nature of correlations decay is an important subject in
 different fields of physics. In particular, after the
 discovery of the so-called dynamical chaos, a large effort
 has been devoted to understand their behavior in
 relation to dynamical properties. The main reason is to know
 the precise conditions under which a statistical
 description is legitimate and to estimate the nature of the
 approximations which are involved.

 Another important characteristic of dynamical systems is the
 stability of their solutions under slight variation of
 the Hamiltonian. A quantitative measure of this stability is
 given by the so-called fidelity or quantum Loschmidt echo.
 The fidelity $ M(t) = |m(t) |^2 $, measures the overlap of two
 states started from the same initial state and
 evolved under slightly different Hamiltonians  $H_0$ and
 $ H=H_0 + \epsilon V $, which are classically chaotic,
 \be m(t) = \la \Psi_0|{\rm exp}(iHt/ \hbar ) {\rm exp}(-iH_0t / \hbar) |\Psi_0 \ra .
 \label{mat} \ee

 Quite surprisingly, in spite of its physical relevance, the behavior
 of fidelity has been scarcely considered and
 only recently, in connection with quantum computation, a large
 number of papers appeared. Some
 important features of fidelity are now understood even though we
 are still far from the detailed level of
 knowledge we have about related quantities such as correlations
 functions and escape probabilities. So far, above
 the perturbative regime of small $\epsilon$,
 with Gaussian type decay \cite{Peres84,Prosen02,CT02},
 two main types of exponential decay of the fidelity have been identified:
 i)- The Fermi-golden-rule (FGR) decay,
 with the exponent given by the half-width of the
 corresponding local spectral density of states\cite{CT02,JSB01,JAB02,EWLC02}.
 This decay has been related to the decay of autocorrelation function \cite{WC02}.
 ii)- The Lyapunov regime,  above the FGR regime, with decay rate given by the Lyapunov
 exponent of the underlying classical dynamics
 \cite{JSB01,JP01,CPW02,WC02,Wis02,CLMPV02,BC02,pre02-fid-kt,VH03,Cucchi03}.

 In this paper we show that for classically chaotic systems, in particular those with
 weak chaos, the behavior of fidelity
 can be much more rich
 and complex than expected. In particular we study perturbation borders
 which separate different types of decay.

%% Figure 1
\begin{figure}
\includegraphics[width=\columnwidth]{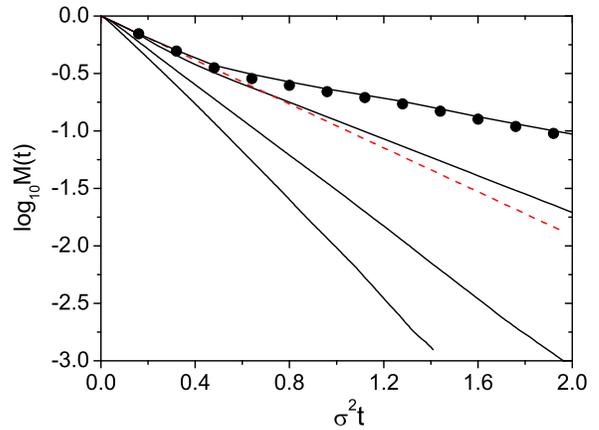}
 \caption{
 Fidelity $\overline M(t)$ as a function of $\sigma^2 t$ for $K_0=0.4$,
 $\epsilon \approx 7.67 \times 10^{-5}$ and
 $N=N_0,2N_0,4N_0,8N_0$ (from bottom to top) where $N_0=4096$.
($\sigma =0.05,0.1,0.2,0.4$).
 The FGR decay $\simeq e^{-2.2\sigma^2 t}$ is shown by the dashed line.
 Full circles represent the semiclassical values $\overline M_a(t)$ at $\sigma =0.4$,
 computed with expression (\ref{Mp-ps}).
 The numerically computed semiclassical values $\overline M_f(t)$ turn out to be
 negligible so that $\overline M(t)$ is well approximated by $\overline M_a(t)$,
 as clearly seen from the figure at $\sigma=0.4$.
 Averages were performed over 400 initial point sources, with $\theta_0$ taken randomly
 in the interval $[0,2\pi )$.
 (The same decaying behaviors are observed for initial Gaussian wavepackets.)}
  \label{fig1}
 \end{figure}

 We start by displaying numerical results which strongly deviate from the expected behavior.
 We consider here the
 simple, well known, sawtooth map model\cite{BC02}. The classical map writes:
 \be \ole p = p +  K_0 (\theta - \pi ), \  \ole {\theta } = \theta + \ole p . \ \
  ({\rm mod} \ 2\pi ) \ee
 For $K_0>0$, the motion is completely chaotic,
 with Lyapunov exponent $\lambda = ln \{ (2+ K_0 +
 [ ( 2+K_0)^2 -4 ]^{1/2} ) /2 \}$.
 The quantum evolution on one map iteration is described by
 \be \label{psi-sa} \ole{\psi } = U_0 \psi \equiv {\rm exp} \left [ - i
 {\hat{p}}^2 /(2 \hbar ) \right ]
 {\rm exp} [ik_0 (\hat{\theta} -\pi )^2/2 ] \psi , \ee
 where $\hat{p} = -i \hbar \partial / \partial \theta $ and
 $k_0 = K_0 / \hbar $,
 with the effective Planck constant $\hbar = 2\pi /N$
 and $N$ being the dimension of the Hilbert space.
 For the perturbed system, $K =K_0 + \epsilon $ and
 $k=k_0 + \sigma $,  where $\sigma=\epsilon/ \hbar $
 and $\epsilon << K_0$.

\begin{figure}
\includegraphics[width=\columnwidth]{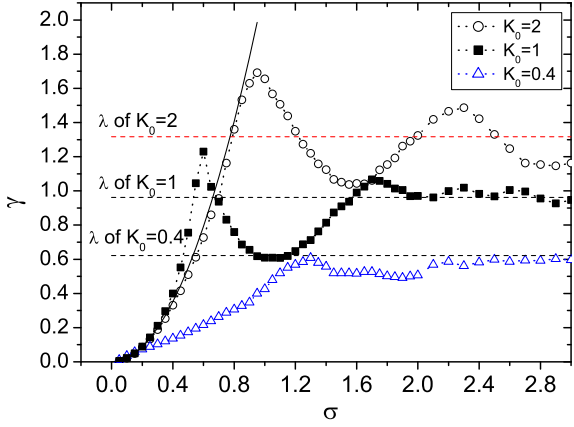}
 \caption{
 The exponential decay rate $\gamma $ versus perturbation
 strength $\sigma $, calculated from the best fit of $\ln \overline M(t)$.
 Gaussian wavepackets are taken as initial states.
 The solid curve shows the rate $\Gamma \simeq 2.2\sigma^2$ of the FGR-decay.
 The dashed horizonal lines correspond to the Lyapunov exponents $\lambda
 =0.62, 0.96$ and 1.32 for $K_0=0.4,1$ and 2, respectively. $N=131072$.
 } \label{fig2}
\end{figure}

 In Fig.~\ref{fig1} we show the fidelity decay in the expected FGR regime $1/\sqrt{N} \lesssim \sigma \lesssim 1$.
 In spite of the fact that
 the classical motion is chaotic, it is clearly seen that the behavior does not obey the FGR which, according to
 \cite{JSB01,BC02}, should be $\propto exp(-\Gamma t)$ with $\Gamma \approx 2.2 \sigma^2$. The same conclusion can
 be drawn from Fig.~\ref{fig2} where we plot the decay rate $\gamma$
 of the fidelity as a function of $\sigma$. Indeed at $K_0 = 0.4$ the
 decay rate $\gamma$ versus $\sigma$ appears quite different from the quadratic one~\cite{note}.

%Fig.2

 Deviations are present even at $K_0=1$ and only at $K_0=2$ one has good FGR decay.
 Moreover, above the FGR regime,
 where one expects Lyapunov decay, there are strong oscillations above and below the decay rate
 $\lambda $ (for $K_0=1$, and $2$).
 Only at larger $\sigma$ values,
 one enters the Lyapunov regime.

 In order to explain the above numerical results, we start from the standard semiclassical approach~\cite{JP01,VH03}.
 For simplicity, we consider a finite configuration space,
 with dimension $d$ and volume  ${\cal V}=\int d\br $.
 The momentum space is also finite, with a volume ${\cal V}_p$.
 In the semiclassical approach, an initial state $\psi_0 (\br_0)$
 is propagated by the semiclassical Van Vleck-Gutzwiller
 propagator,
 $ \psi (\br ; t)= \int d\br_0  K_{\rm sc}(\br , \br_0  ; t)
 \psi_0 (\br_0 )$, where
 $ K_{\rm sc}(\br , \br_0  ; t)= \sum_{s} K_s(\br , \br_0  ; t)$,
 with
 \be K_s(\br , \br_0  ; t) = \frac{C_s^{1/2} } {(2 \pi i \hbar )^{d/2}}
 {\rm exp } \left [ \frac {i}{\hbar } S_s(\br , \br_0  ; t) -
 \frac{i \pi }2 \mu_s \right ]. \label{Ks-exp} \ee
 The label $s$ in Eq.~(\ref{Ks-exp}) (more exactly $s(\br , \br_0 ; t)$),
 indicate classical trajectories
 starting at $\br_0 $ and ending at $\br $ in a time $t$;
 $S_s(\br , \br_0  ; t)$ is the time integral of the Lagrangian
 along the trajectory $s$,
 $ S_s(\br , \br_0  ; t)= \int_0^t dt' {\cal L }$,
 $ C_s = | {\rm det}( \partial^2 S_s / \partial {r_{0i}} \partial r_j )| $,
 and $\mu_s$ is the Maslov index counting the conjugate points.

 In Ref.~\cite{VH03},
 it is shown that the semiclassical approximation to $m(t)$ for initial Gaussian wavepackets
 has a simple and convenient expression in the initial momentum space.
 Following similar arguments for initial point sources,
 $\la \br | \Psi_0  \ra = \sqrt{ (2 \pi \hbar )^d / {\cal V}_p } \delta (\br -\br_0) $,
 ( the theory can be extended to general initial states), one can write $m(t)$ as
 \be  m(\br_0,t)  \simeq  \frac 1{{\cal V}_p}  \int  d{\bf p}_0
 {\rm exp} \left [ \frac i{\hbar} \Delta S({\bf p}_0, \br_0 ; t)  \right ],
 \label{mt-point-p0} \ee
 where $ \Delta S(\bp_0 , \br_0 ; t)$ is the action difference
 along the trajectory starting at $(\br_0,{\bf p}_0)$ for the two systems
 $H_0$ and $H$.
 In the first order classical perturbation theory,
 $ \Delta S(\bp_0 , \br_0 ; t) = \epsilon  \int_0^t dt' V[{\bf r}(t')]$,
 with $V$ evaluated along the trajectory.

    The averaged (over $\br_0$) fidelity can be separated into a mean-value part and a
 fluctuating part \cite{CLMPV02}, denoted by  $\overline M_a(t) $ and $\overline M_f(t)$ respectively,
 $\overline M(t) \equiv \overline{|m(t)|^2} = \overline M_a(t) + \overline M_f(t)$,
 where
 \be \label{Mp} \overline M_a(t) \equiv |  \overline m(t) |^2, \ \ {\rm with} \ \
 \overline m(t) =  \frac 1{\cal V} \int d\br_0 m(\br_0,t). \ee
 From Eqs.~(\ref{mt-point-p0}) and (\ref{Mp}), it is seen that
 the mean-value part $\overline M_a(t)$ can be expressed in
 terms of the distribution $P(\Delta S)$ of the action difference $\Delta S$,
 \bey \overline M_a(t) \simeq \left | \int d\Delta S e^{i\Delta S/ \hbar }
 P(\Delta S)\right |^2, \hspace{0.2cm} {\rm where} \hspace{1.4cm}  \label{Mp-ps}
 \\ P(\Delta S) = \frac 1{ \int d\br_0 d{\bf p}_0 }
 \int d\br_0 d{\bf p}_0 \delta \left [\Delta S - \Delta S(\bp_0 , \br_0;t) \right ].
 \label{PdS}  \eey

  It is usually assumed that for chaotic systems $P(\Delta S)$ is close to a Gaussian
 with a variance $[2 \epsilon^2 K(E)t] $,
 where $K(E) = \int_0^{\infty } dt \la V[\br (t)] V[(\br (0)] \ra $
 is the classical action diffusion constant \cite{CT02}.
 As a result, $\overline M_a(t)\simeq e^{-\Gamma t}$,
 where $\Gamma = 2\sigma^2 K(E) $.
 At small $\sigma $, the fluctuation is small compared with the average value,
 because the phase on the right hand side of Eq.~(\ref{mt-point-p0}) is proportional to $\sigma$;
 then, $\overline M(t) \simeq \overline M_a(t)$ has the FGR-decay.
% Notice that, at zero perturbation $\epsilon =0$, the right hand side of Eq.~(\ref{Mp-ps})
% gives one for perfect fidelity.

 Let us now consider a fixed $\br_0$, and divide
 the space of the initial momenta ${\bf p}_0$
 into connected, disjoint  subspaces, denoted by ${\cal A}_{\alpha }$,
 where each ${\cal A}_{\alpha }$ is the largest possible subspace
 such that the correspondence between $\bp_0$ and the final position $\br $
 is one-to-one, i.e.
 different $\bp_0$ inside each single component ${\cal A}_{\alpha }$ gives different final position $\br$.
 It is always possible to make such a division.
 The number of subspaces ${\cal A}_{\alpha }$ is denoted by $N_{\alpha }$.
 Note also that the sizes of ${\cal A}_{\alpha }$
 decrease exponentially with increasing time $t$.
 When $\bp_0$ runs over a subspace ${\cal A}_{\alpha }$, $\br$ may run over
 part of the configuration space, denoted by ${\cal V}_{\alpha }$.
 Note that, with this division of the ${\bf p}_0$ subspace, the
 trajectories starting at $\br_0$ are divided into $N_{\alpha }$ groups
 and ``near'' trajectories typically belong to the same group.

%Fig.3
\begin{figure}
\includegraphics[width=\columnwidth]{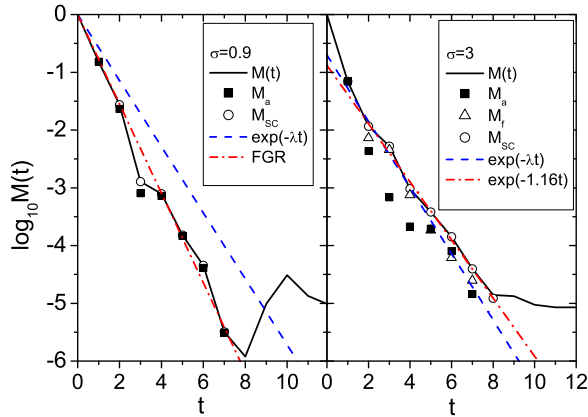}
 \vspace{0.0cm} %\narrowtext
 \caption{
 Comparison between the exact $\overline M(t)$, its semiclassical mean-value part
 $\overline M_a$, calculated by using Eqs.~(\ref{mt-point-p0}) and (\ref{Mp}),
 and the fluctuation part $\overline M_f = \overline M_{\rm sc} - \overline M_a $,
 where $\overline M_{\rm sc}$ is the semiclassical approximation to the fidelity,
 computed by the expression (\ref{mt-point-p0}).
 Here $N=131072$, $K_0=2$, and $\sigma =0.9$ (left panel), $\sigma = 3$(right panel).
 The exact $\overline M(t)$
 is in good agreement with its semiclassical approximation $\overline M_{\rm sc}(t)$.
 The average is taken over 500 initial point sources.
  } \label{fig3}
 \vspace{0.8cm} %\narrowtext
\end{figure}

    The amplitude $m(\br_0,t)$  in Eq.~(\ref{mt-point-p0}) can now be written as
 $m(\br_0,t) \simeq \sum_{\alpha } m_{\alpha }(\br_0,t)$, where
 \be m_{\alpha }(\br_0,t)  =  \frac{1}{{\cal V}_p }  \int_{ {\cal V}_{\alpha}}  d\br
 C_{ s}  {\rm exp} \left [ \frac i{\hbar} \Delta S_s(\br , \br_0 ; t)  \right ]
 \label{mt-alpha} \ee
 with integration over the subspace ${\cal V}_{\alpha }$,
 in which the change of variable $\bp_0 \to \br$ within the subspaces ${\cal A}_{\alpha }$
 has been done
 and $\Delta S_s(\br , \br_0 ; t)$ coincides with $\Delta S(\bp_0 , \br_0 ; t)$
 for the same trajectory $s$ starting at $(\br_0, \bp_0)$
 with $\bp_0 \in {\cal A}_{\alpha }$.
 $\overline M_f(t)$ is written as
 \be \label{Mf-malpha}  \overline M_f(t)
 \simeq  \overline { \left | \sum_{\alpha } m_{\alpha f} \right |^2 }
 \hspace{0.1cm} {\rm with } \
 m_{\alpha f} = m_{\alpha }(\br_0,t) - \frac{\overline m(t)}{N_{\alpha }}. \ee
 When $\sigma $ is large enough, above a critical border $\sigma_f$,
 $m_{\alpha }(\br_0,t)$ can be regarded as possessing random phase, and therefore
 $\overline M_f$ can be approximated by its diagonal part
 \bey \overline M_f(t) \simeq  \overline { \sum_{\alpha}  |m_{\alpha f}|^2 }
 \simeq  \overline { \sum_{\alpha}  |m_{\alpha }(\br_0,t)|^2 }
 \hspace{2cm} \nonumber
 \\ \propto \int d \br_0 \sum_{\alpha }
 \left | \int_{{\cal V}_{\alpha }}  d\br
 C_{ s} {\rm exp} \left ( \frac i{\hbar} \Delta S_s \right )
 \right |^2, \label{ma-sa} \eey
 where the second approximation is obtained by noticing that
 $|\overline m(t) / N_{\alpha }| << |m_{\alpha }|$ at large $\sigma $.

      When the phase space is homogeneous with constant local
 (maximum) Lyapunov exponent $\lambda $, as in the sawtooth map,
 the number $N_s(\br_0,\br )$ of trajectories connecting two points
 $\br_0$ and $\br $ in the configuration space is approximately
 $ N_s\simeq N_{\alpha }\simeq e^{\lambda t}$\cite{Haake}.
 The summation over $\alpha $ in (\ref{ma-sa})
 gives a contribution approximately proportional to $N_s$.
 At $t$ large enough, the main time dependence of
 $ |\int_{{\cal V}_{\alpha }} d\br C_s e^{ \frac i{\hbar} \Delta S_s }|$
 is given by $C_s \propto e^{-\lambda t}$.
 Combining these results, it is seen that at $\sigma > \sigma_f$,
 $\overline M_f(t)$ has the Lyapunov decay,
 $\overline M_f(t) \propto e^{-\lambda t}$.

%Fig.4
\begin{figure}
\includegraphics[width=\columnwidth]{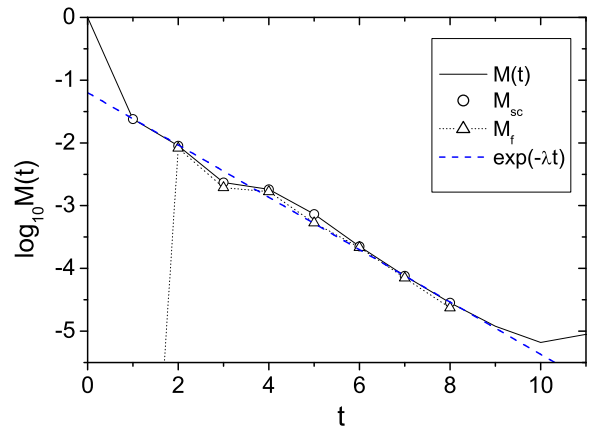}
 \caption{
 Similar to Fig.~\ref{fig3}, for $K_0=1$ and $\sigma =6$.
 At this large $\sigma $,
 $\overline M_a(t)$ is negligible compared with $\overline M_f(t)$.
 } \label{fig4}
 \vspace{0.8cm} %\narrowtext
\end{figure}

 In order to have the Lyapunov decay for  $\overline M(t)$, the term $\overline M_a(t)$ must be small.
 To this end one needs to further increase $\sigma$ above a critical value $\sigma_r$, so that
 the variance of the phase of $m(\br_0, t)$ with respect to $\br_0$ will become so large that
 $\overline M_a(t)$ is negligible.

 The right panel of Fig.~\ref{fig3} gives an example of $\overline M_a(t) \approx \overline M_f(t)$.
 This
 explains the fluctuation of $\gamma $ versus $\sigma$ shown in Fig.~\ref{fig2} at $K_0=2$ and $\sigma <3$.
 Fig.~\ref{fig4} instead gives an example with
 $\sigma $ large enough ($\sigma > \sigma_r$), so that $\overline M_a(t)$ is negligible and
 $\overline M(t) \simeq \overline M_f(t)$.

%Fig.5
\begin{figure}
\includegraphics[width=\columnwidth]{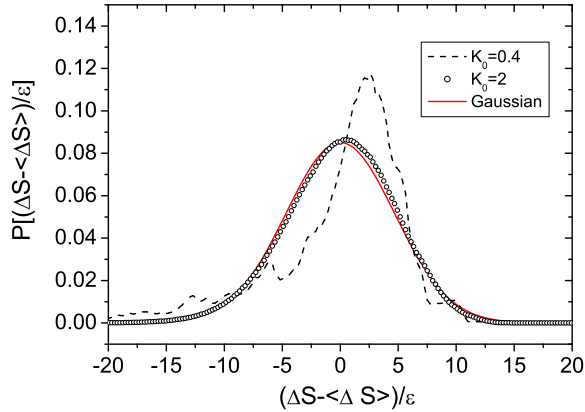}
 \caption{
 Distribution $P[(\Delta S -\la \Delta S \ra )/ \epsilon]$
 of the classical action difference $\Delta S$,  at $t=10$,
 calculated by taking randomly $10^7$ initial points in the phase space,
 where $\la \Delta S \ra \equiv \epsilon t \la V(\theta ) \ra
 = -\pi^2 \epsilon t/6$, with an average over the phase space.
 } \label{fig5}
\end{figure}

 The deviation from FGR decay observed in Figs.~\ref{fig1} and \ref{fig2}
 is due to the deviation of $P(\Delta S)$
 from the Gaussian behavior.
 Indeed, when chaos in the underlying classical dynamics is strong enough
 ($K_0 > 1$), correlations between
 non-overlapping parts of a trajectory
 decay very rapidly and the distribution $P(\Delta S / \epsilon)$ reaches, in a relatively short time,
 the Gaussian distribution.
 This is the case of Fig.~\ref{fig5} for $K_0=2$ , where $K(E) = \pi^4/90 \simeq 1.08$
 with $\Gamma = 2K(E) \sigma^2 \simeq 2.16 \sigma^2$,
 in agreement with the numerical results in Ref.~\cite{BC02}
 and in Fig.~\ref{fig2}.
 However, when $K_0$ is not sufficiently large, e.g., $K_0=0.4$,
 a considerable deviation of $P(\Delta S / \epsilon)$  from the Gaussian distribution
 appears for times comparable to the fidelity decay times (Fig.~\ref{fig5}).
 According to Eq.~(\ref{Mp-ps}),
 this leads to deviations from the FGR-decay as observed in Fig.~\ref{fig1}.
 We would like to draw the reader attention to the fact that, for $K_0<1$ the saw-tooth map, even though
 completely chaotic, it possess a structure of cantori which, in the quantum case, can act as perfect barriers
 to quantum motion thus leading to localization of wavefunctions.

 Notice that the deviation of $P(\Delta S / \epsilon)$  from the Gaussian distribution
 depends on $K_0$ but not on $\epsilon$  or $\sigma$. Therefore, by increasing $\sigma$, the effect of
 this deviation becomes more and more important,
 since the FGR exponential decay has a decay rate
 proportional to $\sigma^2$ while the deviation from the Gaussian remains unchanged.
 Therefore, for a given system, there is a critical value $\sigma_d$, below which the FGR decay is
obeyed with good accuracy
 and above which FGR breaks down.
 This case is illustrated in Fig.~\ref{fig2} for the case $K_0=1$, which coincides with
 the well-known Arnold cat map,
 the paradigmatic model of chaos. Here
 the distribution $P(\Delta S / \epsilon )$ (at t= 10) is slightly different from the Gaussian
 distribution and
 the decay rate $\gamma $ of fidelity deviates from the FGR decay for $\sigma \gtrsim  0.3$.
 In cases of weak classical chaos, the
 value of $\sigma_d$ can be so small that FGR is never observed (e.g. the case with  $K_0=0.4$).
 The left panel in Fig.~\ref{fig3} shows instead a case at
 $K_0=2$ and $\sigma =0.9$, in which  $\overline M_a(t)$ obeys
the FGR decay and $\overline M_f(t)$ is negligible.

 To summarize: above the perturbative border, the fidelity has a
 FGR decay for $\sigma < \sigma_d$,
 while for $\sigma > \sigma_r$, it has the Lyapunov decay.
 In the intermediate region, for $\sigma_d < \sigma < \sigma_f$,
 the fidelity deviates
 from FGR and can decay even faster than Lyapunov.
 For $\sigma_f < \sigma < \sigma_r$,
 $\overline M_a(t) \sim \overline M_f(t)$
 and the decay rate of $\overline M(t)$
 fluctuates around the Lyapunov exponent.
 It may be useful to recall here the physical meaning of different borders.
 Above $\sigma_d$, the distribution
 $P(\Delta S )$ deviates
 from the Gaussian and this induces deviations from the expected FGR decay.
 Below $\sigma_r$, $\overline M_a(t)$ is non negligible as
 compared to $\overline M_f(t)$ and this induces deviations from the expected Lyapunov decay.

 It may be interesting to remark that the relation between the
 decomposition in two part of $\overline M(t)$ here and that in Ref.~\cite{JP01} is the following.
 At $\sigma $ small enough $\overline M(t) \simeq \overline M_a(t) \simeq M^{nd}(t)$,
 with $\overline M_f(t)$ and $M^d(t)$ negligible;
 while at $\sigma $ large enough $\overline M(t) \simeq \overline M_f(t) \simeq M^d(t)$,
 with $\overline M_a(t)$ and $ M^{nd}(t)$ negligible.
 In the intermediate regime of $\sigma $,
 in particular, in the crossover from the FGR decay to the Lyapunov decay,
 there may be  considerable difference between the two divisions.

 In this paper, by using the sawtooth map, we have demonstrated that
 the fidelity decay in a generic chaotic system
 can have a very complex behavior. In particular,
 deviations from the Fermi golden rule (for weak chaos) and
 Lyapunov decay have been discussed as well as the existence of
 perturbation borders separating different regimes.
 It is our opinion that fidelity is an important
 quantity which characterizes the stability of classical and
 quantum systems. It therefore deserves deeper analytical and
 numerical studies in order to fully understand its
 behavior in different dynamical regimes.

 The authors are grateful to V.~Sokolov for valuable discussions.
 This work was supported in part by the Academic Research
 Fund of the National University of Singapore.
 Support was also given by
 the EC RTN contract HPRN-CT-2000-0156,
 the NSA and ARDA under ARO contracts No.
 DAAD19-02-1-0086, the project EDIQIP of the IST-FET programme of
 the EC, the PRIN-2002 ``Fault tolerance,
 control and stability in quantum information processing'',
 and the Natural Science Foundation of China No.10275011.


\begin{thebibliography}{99}
\bibitem{note} Deviations from Fermi golden rule have also been observed
 in the Bunimovich stadium billiard~\cite{Wis02}.
 The fidelity in that model was found to decay as ${\rm exp} (-\Gamma t / 2)$,
 with $\Gamma $ being the half width of the LDOS.
 In our case the decay rate of fidelity is close to $\Gamma$.
 However, for $K_0 =0.4$ considerable deviations in tail region of LDOS from the Lorentzian shape are present.
\end{thebibliography}
 \end{document}